\documentclass[journal, comsoc]{IEEEtran}

\usepackage{cite}

\ifCLASSINFOpdf
   \usepackage[pdftex]{graphicx}
   \graphicspath{{./images/}}
   \DeclareGraphicsExtensions{.pdf,.jpeg,.png}
\else
   \usepackage[dvips]{graphicx}
   \graphicspath{{./images/}}
   \DeclareGraphicsExtensions{.eps}
\fi

\usepackage{amsmath}
\usepackage{amssymb}
\usepackage[shortlabels, inline]{enumitem}

\usepackage{algorithmic}

\usepackage{array}

\usepackage{tabularx}
\usepackage{booktabs}
\usepackage{multirow}
\usepackage{makecell}
\usepackage[flushleft]{threeparttable}

\ifCLASSOPTIONcompsoc
   \usepackage[caption=false, font=normalsize, labelfont=sf, textfont=sf]{subfig}
\else
   \usepackage[caption=false, font=footnotesize]{subfig}
\fi

\usepackage{dblfloatfix}

\ifCLASSOPTIONcaptionsoff
   \usepackage[nomarkers]{endfloat}
   \let\MYoriglatexcaption\caption
   \renewcommand{\caption}[2][\relax]{\MYoriglatexcaption[#2]{#2}}
\fi

\usepackage{url}

\usepackage{xcolor}
\usepackage{nameref}

\newcolumntype{Y}{>{\centering\arraybackslash}X}

\newcommand{\etal}{\textit{et al}. }

\begin{document}

\title{DRL-GAN: A Hybrid Approach for Binary and Multiclass Network Intrusion Detection}

\author{
   \IEEEauthorblockN{
       Caroline~Strickland\IEEEauthorrefmark{1}, Chandrika~Saha\IEEEauthorrefmark{2}, Muhammad~Zakar\IEEEauthorrefmark{3}, \\ Sareh~Soltani~Nejad\IEEEauthorrefmark{4}, Noshin~Tasnim\IEEEauthorrefmark{5},
       Daniel~Lizotte\IEEEauthorrefmark{6}, Anwar~Haque\IEEEauthorrefmark{7}
   }\\%
   \IEEEauthorblockA{
       Department of Computer Science, The University of Western Ontario, London, Canada\\
       \{\IEEEauthorrefmark{1}cstrick4,  \IEEEauthorrefmark{2}csaha, \IEEEauthorrefmark{3}mzakar,  \IEEEauthorrefmark{4}ssolta7, \IEEEauthorrefmark{5}ntasnim3, \IEEEauthorrefmark{6}dlizotte, \IEEEauthorrefmark{7}ahaque32\}@uwo.ca
   }%
}

\maketitle

\begin{abstract}
    Our increasingly connected world continues to face an ever-growing amount of network-based attacks. Intrusion detection systems (IDS) are an essential security technology for detecting these attacks. Although numerous machine learning-based IDS have been proposed for the detection of malicious network traffic, the majority have difficulty properly detecting and classifying the more uncommon attack types. In this paper, we implement a novel hybrid technique using synthetic data produced by a Generative Adversarial Network (GAN) to use as input for training a Deep Reinforcement Learning (DRL) model. Our GAN model is trained with the NSL-KDD dataset for four attack categories as well as normal network flow. Ultimately, our findings demonstrate that training the DRL on specific synthetic datasets can result in better performance in correctly classifying minority classes over training on the true imbalanced dataset.
\end{abstract}


\section{Introduction}
\label{sec:introduction}

\IEEEPARstart{T}{he} increasing volume and sophistication of network-based attacks motivate the development of effective techniques and tools to prevent service disruption, unauthorized access, and the disclosure of sensitive information~\cite{Hsu2020}. An Intrusion Detection System (IDS) is an important defence tool against sophisticated and increasing network attacks, but these systems, especially Machine Learning (ML) based systems, require large, reliable, and valid network traffic datasets to be effective. Although the majority of recently available datasets cover a range of network attack types and traffic patterns and include information about the attacking infrastructure, modern networks are increasingly diversified such that existing datasets are often not enough to develop effective classification mechanisms. These datasets often suffer from a lack of traffic diversity and volume or fail to cover the full scope of known attack types. To cope up with these new changes, we require a more dynamic dataset that will improve the ability of an IDS to detect intrusions. Using deep learning techniques such as Generative Adversarial Networks (GANs), we can fabricate additional data using existing datasets to increase the classification accuracy of an IDS, especially for rare attack categories.

Two methods of IDS are Signature-based Intrusion Detection Systems (SNIDS) and Anomaly-based Intrusion Detection Systems (ANIDS). The SNIDS approach is effective for known threats, as it looks for specific patterns (or `signatures') such as byte sequences in network traffic, or known malicious instructions sequences used by malware~\cite{Hsu2020}. Conversely, the ANIDS approach uses ML algorithms to analyze and monitor the network traffic in order to detect any suspicious activity, thus being an effective method for catching unknown attacks~\cite{bhuyan2013network}.

The emergence of deep learning and its integration with Reinforcement Learning (RL) has created a class of Deep Reinforcement Learning (DRL) methods that are able to detect the most recent and sophisticated types of network attacks. DRL combines artificial neural networks with a framework of RL that helps software agents (or `learning entities') learn how to reach their goals. DRL combines function approximation and target optimization, mapping states and actions to the rewards they lead to~\cite{li2017deep}. This results in a `policy' that our learning agents can follow to make the best decisions given the current state. To detect network attacks, DRL is used to train an agent such that, given a `state' represented as a collection of feature values, will take the best `action' (which, in our case, acts as a classification of attack type), in order to recognize an attack.

Each network is different in that its behaviours and patterns evolve gradually. Naturally, vulnerabilities also evolve. The performance of IDS classification accuracy suffers as existing datasets gradually become out of date, invalid, and unreliable. Moreover, reliable data cannot often be shared due to privacy concerns. Existing publicly available datasets do not include all of the existing network attack types, let alone the unknown vulnerabilities and attacks. To resolve this, we need more diverse and up-to-date datasets that properly reflect the characteristics of network intrusions in order to increase the performance of the IDS. Knowing this, we propose a SNIDS using DRL techniques. We use a collection of GAN models to generate varied datasets, then use DRL to implement an IDS and train the model on the GAN-generated datasets and compare our results.

We use the open-source dataset \mbox{NSL-KDD}~\cite{Tavallaee2009}. NSL-KDD is imbalanced with significantly less attack samples than normal traffic (especially for Probe, U2R, and R2L attacks). Thus, we used GAN to generate synthetic data so that there is a more even class balance. We then trained the DRL model on both the untouched NSL-KDD dataset as well as the GAN-generated data from each of our unique models for both binary and multiclass classification. Finally, we assess how training the DRL models using synthetic datasets compares in terms of IDS performance as well as individual class F1-scores.

Overall, the primary contributions of this paper \mbox{include}:
\begin{enumerate}
    \item Using both conditional and unconditional CTGAN and copulaGAN models to generate tabular data. This is useful for increasing the minority class samples in imbalanced datasets, as well as providing large datasets for training ML models.
    \item Combining GAN and DRL techniques for the purpose of network intrusion detection and increasing the precision and recall for classifying underrepresented class data. We propose a framework that trains a GAN model to produce synthetic data, and then passes that data to a DRL model that acts as an IDS and either alerts a user to an attack or classifies the network traffic as benign.
\end{enumerate}

The remainder of this paper is organized as follows: Section~\ref{sec:related_work} surveys related work for the purpose of network intrusion detection and presents the motivation and novelty behind this work. Section~\ref{sec:methodology} discusses methodology and details necessary for implementation of our models. Section~\ref{sec:results} provides a comprehensive evaluation of the obtained results. Section~\ref{sec:conclusion} presents an interpretation of our findings. Finally, Section~\ref{sec:future_work} discusses directions for future work.

\section{Related Work}
\label{sec:related_work}

Hsu and Matsuoka~\cite{Hsu2020} propose a DRL model for anomaly-based network intrusion detection. 
This approach treats the network traffic data as the RL environment state variables and the outcome of intrusion detection as the action. The correctness of the intrusion recognition result is used to determine the reward. The novelty of this work is that the DRL agent dynamically alternates between `detection mode' and `learning mode' based on whether the current performance of the system is below a predefined threshold. 
In learning mode, the performance is evaluated through the reward and the model is updated with the new traffic data to improve the detection performance. In detection mode, a dummy reward is used to maintain operation and the true reward of the label is not calculated. The system was evaluated on pre-established benchmark datasets, \mbox{NSL-KDD} \cite{Tavallaee2009} and \mbox{UNSW-NB15}~\cite{Moustafa2016}, and consistently achieved over 90\% in  accuracy, recall, and precision performance metrics. 

Alavizadeh~\etal \cite{alavizadeh2022deep} also propose a DRL-based continuously updating, self-learning NIDS. Their proposed Deep Q-Learning (DQL) model combines Q-learning based RL with a deep feed forward neural network to detect network intrusions. The model uses an ongoing trial and error auto-learning approach to improve its detection capabilities for different types of network intrusions. The model was evaluated on the \mbox{NSL-KDD}~\cite{Tavallaee2009} dataset and outperformed some other ML techniques with 78\% classification accuracy for the intrusion classes. This work is promising similarly to work in~\cite{Hsu2020} due to its adaptive-learning capabilities, making it better suited for securing networks from the inevitably more sophisticated attacks cyber-attacks seen today.

Benaddi~\etal \cite{Benaddi2020} developed a DRL-based IDS (\mbox{DRL-IDS}) for Wireless Sensor Networks (WSNs)~\cite{Patil2017} and Internet of Things (IoTs)~\cite{Kumar2019}. Networking architectures like WSNs and IoTs are receiving increasingly more adoption in many areas such as healthcare, business, and smart cities and cyber-threats are the primary challenge for these networks~\cite{Ghosh2019}. They highlight that these networks are vulnerable to intrusions due to security flaws commonly found in IoT and WSN devices, zero-day vulnerabilities, and the openness of these networks to a large number of users. The DRL-IDS model improves intrusion detection performance while monitoring real-time network traffic. The model was evaluated against standard RL and K-Nearest Neighbours (KNN) based approaches using the \mbox{NSL-KDD}~\cite{Tavallaee2009} dataset and performed better in terms of accuracy and detection rate with fewer false negatives.

Lin~\etal~\cite{Lin2018} propose a IDSGAN, a framework that uses GANs to generate adversarial malicious network traffic to deceive IDS. Their goal was to leverage GANs to improve IDS by exposing them to new, more combative and adversarial attack methods and types. This system modeled the black-box analogy of IDS from the perspective of an attacker that would generally not know about the internal details of the detection system. A generator transformed known malicious traffic records into adversarial ones and a discriminator classified the records to learn about the originally unknown detection system. The authors demonstrated the validity of their system by only modifying the nonfunctional features of the records such that the modified records would still classify as an intrusion and not junk traffic. They evaluated their system using the standard \mbox{NSL-KDD}~\cite{Tavallaee2009} dataset on multiple different detection models including Naive Bayes, Random Forest, and multilayer perceptron classifiers. IDSGAN achieved excellent results. The detection rate of the DoS attack type dropped from approximately 80\% with normal records to less than 1\% with modified, adversarial records.

Ring~\etal~\cite{Ring2018} used GANs to generate realistic flow-based network traffic data. They highlight that the ability of GANs to only process continuous attributes is a key challenge in using GANs to generate network traffic since network traffic data ultimately contains categorical features like IP addresses and ports. They propose three preprocessing techniques for converting categorical values in flow-based network traffic data into continuous values:
\begin{enumerate*}[(1)]
    \item Simply treat features such as IP addresses and ports as numerical values
    \item Create binary features from the categorical features
    \item Use IP2Vec~\cite{Ring2017-b} to represent the categorical features as vectors.
\end{enumerate*}
The authors evaluated these techniques on the \mbox{CIDDS-001}~\cite{Ring2017-a} dataset and found that techniques (2) and (3) are effective at generating high-quality flow-based network traffic data. Finally, technique (1) is not well suited for this task, meaning that straightforward numeric interpretation of categorical features should be avoided with GANs.

Overall, there have been a handful of studies focused on using DRL to classify network traffic as normal or intrusion, as well as several that have used GANs to generate network traffic data. However, no study has combined these two ML approaches and evaluated the viability and effectiveness of this combination both in detecting and classifying network traffic as well as increasing the precision and recall performance for classifying previously underrepresented classes. Our proposed solution bridges this gap and improves the current state of knowledge in this field.

\section{Methodology}
\label{sec:methodology}

\subsection{NSL-KDD Dataset}
\label{nsl_dataset}

\mbox{NSL-KDD} is an updated version of the KDD’99 dataset~\cite{Tavallaee2009}. Basic processing has been done, such as the removal of redundant records preventing classifiers from becoming biased towards more frequent records. The use of the NSL-KDD dataset has been very popular in studies on IDS, in a sense, becoming the de facto standard. It contains information which can help to build a host-based and network-based intrusion detection model to ensure network security in a variety of systems.

The training and test set contains 125\,973 and 22\,544 records, respectively. This includes 42 features, however we remove `Num outbound cmds' as all records contain 0, so we are left with 41 features: 9 basic features, 12 content features for the connection, 9 temporal features calculated at two-second time windows, 10 statistical network traffic features, and the class label. Table~\ref{tab:nsl_kdd_features} lists the features present in the dataset. The training set contains 22 attack types and the test set contains 37 attacks types. The 15 attack types not included in the training set make this dataset excellent for modelling unknown attacks. We opt to use the common 5-class classification of network traffic records: normal, DoS, Probe, R2L, and U2R. Table~\ref{tab:nsl_kdd_classes} describes these 5 classes in further detail. The class ID refers to the numerical mapping used by DRL and GAN.

\begin{table}[t]
    \begin{threeparttable}
        \centering
        \caption{NSL-KDD Dataset Features}
        \label{tab:nsl_kdd_features}
        \begin{tabularx}{\columnwidth}{m{0.05\columnwidth}Xm{0.05\columnwidth}X}
            \toprule
            \textbf{F\#} & \textbf{Feature} & \textbf{F\#} & \textbf{Feature} \\
            \midrule
            {F1} & {Duration} & {F22} & {Is guest login} \\
            F2 & Protocol\_type & F23 & Count \\
            F3 & Service & {F24} & {Srv count} \\
            F4 & Flag & F25 & Serror rate \\
            {F5} & {Src bytes} & F26 & Srv serror rate \\
            {F6} & {Dst bytes} & F27 & Rerror rate \\
            {F7} & {Land} & F28 & Srv rerror rate \\
            {F8} & {Wrong fragment} & F29 & Same srv rate \\
            {F9} & {Urgent} & {F30} & {Diff srv rate} \\
            {F10} & {Hot} & {F31} & {Srv diff host rate} \\
            {F11} & {Num\_failed\_logins} & F32 & Dst host count \\
            F12 & Logged\_in & F33 & Dst host srv count \\
            {F13} & {Num compromised} & F34 & Dst host same srv rate \\
            {F14} & {Root shell} & {F35} & {Dst host diff srv rate} \\
            {F15} & {Su attempted} & {F36} & {Dst host same src port rate} \\
            {F16} & {Num root} & {F37} & {Dst host srv diff host rate} \\
            {F17} & {Num file creations} & F38 & Dst host serror rate \\
            {F18} & {Num shells} & F39 & Dst host srv serror rate \\
            {F19} & {Num access files} & F40 & Dst host rerror rate \\
            {F20*} & {Num outbound cmds} & F41 & Dst host srv rerror rate \\
            {F21} & {Is host login} & F42 & Class label \\
            \bottomrule
        \end{tabularx}
        \begin{tablenotes}
            \scriptsize
            \setlength\labelsep{0pt}
            \item * Removed during data preprocessing.
        \end{tablenotes}
    \end{threeparttable}
\end{table}

\begin{table}[t]
    \renewcommand{\arraystretch}{1.3}
    \centering
    \caption{NSL-KDD Dataset Record Classes}
    \label{tab:nsl_kdd_classes}
    \begin{tabularx}{\columnwidth}{cccc>{\raggedright\arraybackslash}X}
        \toprule
        \textbf{ID} & \textbf{Class} & \textbf{Symbol} & \textbf{\# of Records} & \multicolumn{1}{c}{\textbf{Definition}} \\
        \midrule
        0 & Normal & N & 77\,054 & Normal network traffic record \\
        1 & DoS    & D & 53\,387 & Denial of Service attack to prevent requests from intended users from being fulfilled \\
        2 & Probe  & P & 14\,077 & Probing attack to gather information such as vulnerabilities about the target machine or network \\
        3 & R2L    & R & 3880    & An attacker tries to gain local access by sending packets to a remote machine \\
        4 & U2R    & U & 119     & An attacker with normal access tries to gain access to the root by exploiting system vulnerabilities \\
        \bottomrule
    \end{tabularx}
\end{table}



\subsection{Machine Learning Performance Evaluation}

\begin{figure}[t]
    \centering
    \includegraphics[width=0.85\columnwidth]{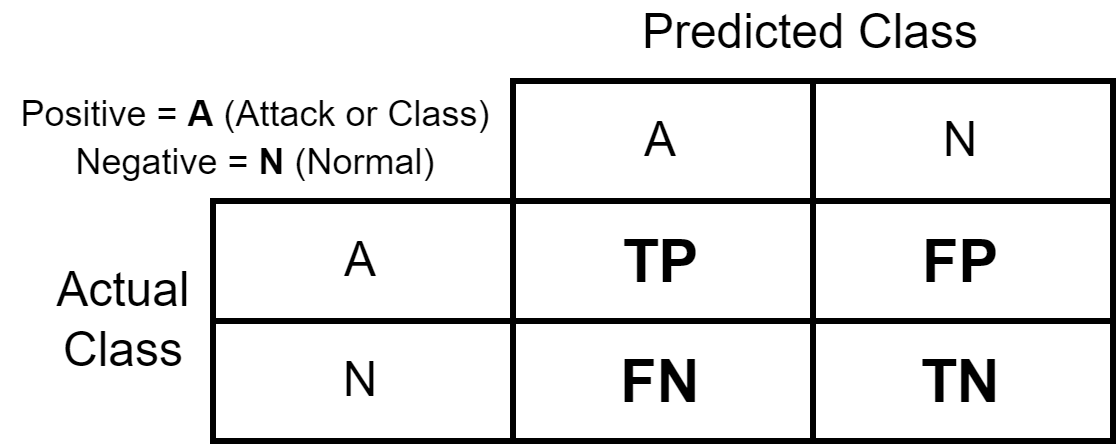}
    \caption{Confusion matrix for NIDS performance evaluation.}
    \label{fig:performance_metrics_cf}
\end{figure}

We used the accuracy and F1-score (which combines precision and recall) metrics to evaluate the performance our DRL model and other ML algorithms. While the accuracy score only measures the percentage of correctly classified samples, this selection of performance metrics allows us to also evaluate the percentage of samples that were incorrectly classified. This is especially important for NIDS as the accuracy performance metric is not enough to evaluate imbalanced datasets such as network traffic data which generally include significantly more normal traffic. These performance metrics are derived from the True Positive (TP), True Negative (TN), False Positive (FP), and False Negative (FN) values. Fig.~\ref{fig:performance_metrics_cf} presents this confusion matrix used by our evaluation method.

\subsubsection{Accuracy}

Accuracy measures the number of correct predictions out of the total predictions made by the model. In this case, accuracy measures the model's ability to correctly identify normal and attack traffic records. Equation~\ref{eqn:accuracy} formalizes the accuracy performance metric.

\begin{equation}
    \label{eqn:accuracy}
    \mathit{Accuracy = \frac{TP + TN}{TP + FP + TN + FN}}
\end{equation}

\subsubsection{Precision}

Precision measures the number of correct positive predictions out of the total number of positive predictions. In this case, precision measures the model's degree of correctness in predicting attack records over the total number of attacks predicted~\cite{Hsu2020,Alghayadh2020-a,Alghayadh2020-b}. Equation~\ref{eqn:precision} formalizes the precision performance metric.

\begin{equation}
    \label{eqn:precision}
    \mathit{Precision = \frac{TP}{TP + FP}}
\end{equation}

\subsubsection{Recall}

Recall measures the number of correct positive predictions out of the total number of positive instances in the dataset. In this case, recall measures the model's ability to correctly identify attack traffic records. From this definition, recall is also referred to as the true positive rate, detection rate, or sensitivity. Equation~\ref{eqn:recall} formalizes the recall performance metric.

\begin{equation}
    \label{eqn:recall}
    \mathit{Recall = \frac{TP}{TP + FN}}
\end{equation}

\subsubsection{F1-Score}

F1-score is the harmonic mean of the precision and recall values, essentially a combined measure of the two performance metrics. F1-score quantifies how discriminative the model is~\cite{Alghayadh2021} and acts as a good indicator of performance since a decrease in either precision or recall results in a significant decrease in the F1-score. In addition, for multiclass classification we present both the unweighted and weighted F1-scores. The weighted F1-score accounts for label imbalance by considering the number of instances of each label when calculating the average F1-score. Equation~\ref{eqn:f1_score} shows how the F1-score is calculated.

\begin{equation}
    \label{eqn:f1_score}
    \begin{aligned}
        \mathit{F1\ Score} &= 2 \cdot \frac{Precision \cdot Recall}{Precision + Recall} \\
        &= \frac{TP}{TP + \frac{1}{2}(FP + FN)}
    \end{aligned}
\end{equation}

\subsection{Statistical Evaluation of Synthetic Data}

To evaluate the synthetic data generated by the GAN models against the real data they were trained on, we used statistical metrics that compare the columns of the synthetic tables against those in the real tables. These statistical metrics are as follows:

\subsubsection{CSTest}

The CSTest compares columns with discrete values using the Chi-squared test to compare their distributions. The output of the test is an average of the CSTest $p$-values for each of the columns, which ultimately quantifies the probability that the compared columns were sampled from the same distribution.

\subsubsection{KSTest}

The KSTest compares columns with continuous values using the two-sample Kolmogorov–Smirnov test and empirical Cumulative Distributed Function (CDF) to compare their distributions. The output of the test is an average of 1 minus the KSTest D statistic for each of the columns. This result quantifies the maximum distance between the CDF expected and observed values.

\subsubsection{KSTestExtended}

The KSTestExtended is an extension of the KSTest that converts all columns to numerical values using a hyper transformer and then applies the regular KSTest.

\subsection{Detection-based Evaluation of Synthetic Data}

Detection metrics use ML models to determine how distinguishable the synthetic data is from the real data. To achieve this, both the synthetic and real tables are shuffled and a flag indicating whether the record is synthetic or not is added. Next, cross-validation is used with a selected ML model that predicts the flag, outputting 1 minus the average ROC AUC of all the cross-validation splits. Because the ROC AUC measures the separability of the classes from the model, a high detection metric score means that the model is unable to easily distinguish the synthetic records from the real ones.

\subsection{Generative Adversarial Network Models}

\begin{figure}[t]
    \centering
    \includegraphics[width=\columnwidth]{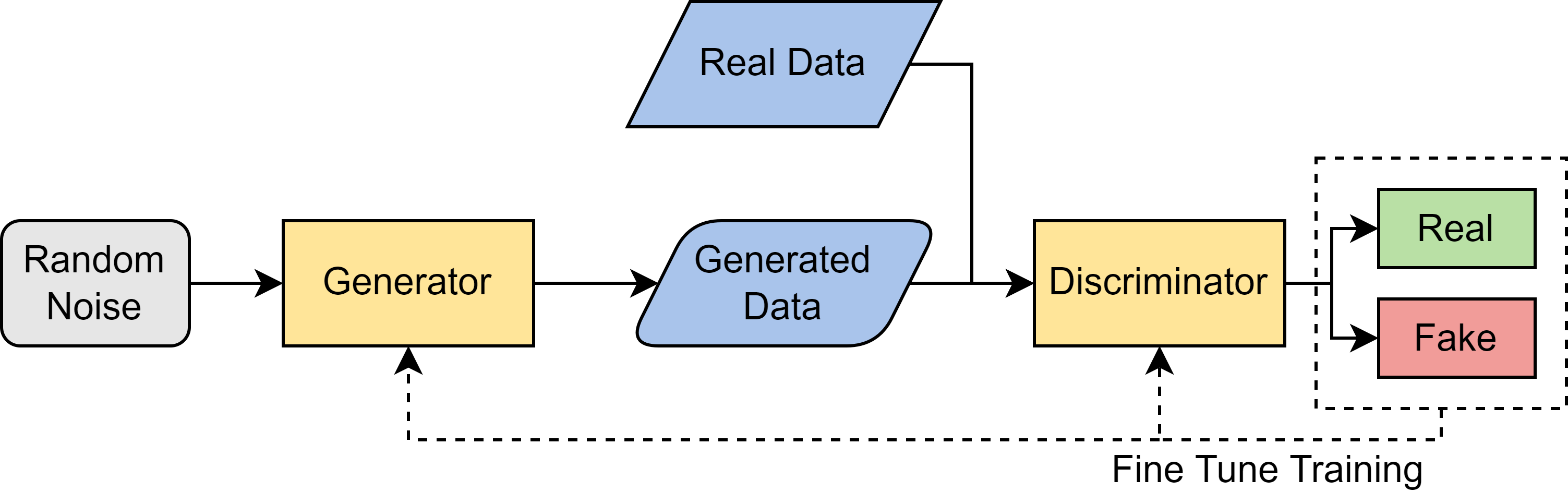}
    \caption{Architecture of Generative Adversarial Networks.}
    \label{fig:gan_architecture}
\end{figure}

Goodfellow~\etal~\cite{gan} first proposed the idea of a GAN in 2014 as an unsupervised learning method that generates synthetic data using an input of real data. GANs are used to generate realistic synthetic data using real data, usually because obtaining more data can be difficult, time consuming, and costly. GANs use two independent models, a generator and a discriminator. By detecting patterns or similarity from given input data, the generator processes input data and produces more data. The discriminator is a classifier which determines the difference between the real data and the generated data. It produces a probability between 0 and 1 to define whether an instance belongs to the real data (closer to 1) or to the generated data (closer to 0). Fig.~\ref{fig:gan_architecture} highlights the overall workflow of GANs.

\subsection{Deep Reinforcement Learning Model}

DRL is a subfield of ML that combines both RL and deep learning. RL considers the problem of an agent learning to make decisions through trial and error, while DRL incorporates deep learning, allowing agents to make decisions from unstructured input data without manual engineering of the state space.

RL problems involve an agent learning how to map situations to actions in order a maximize a numerical reward signal. It employs five key concepts:
\begin{itemize}
    \item \textbf{Environment}: The physical world that the agent operates within.
    \item \textbf{State}: The agent's belief of a configuration of the environment.
    \item \textbf{Reward}: Numerical feedback from the environment.
    \item \textbf{Policy}: A mapping from the agent's state to actions.
    \item \textbf{Value}: Expected future reward an agent would receive by taking an action in a certain state.
\end{itemize}

Simply put, RL is the process of running an agent through sequences of state-actions pairs, observing the rewards that result, and using those rewards to formulate an optimal policy over time.

\begin{figure}[t]
    \centering
    \includegraphics[width=0.9\columnwidth]{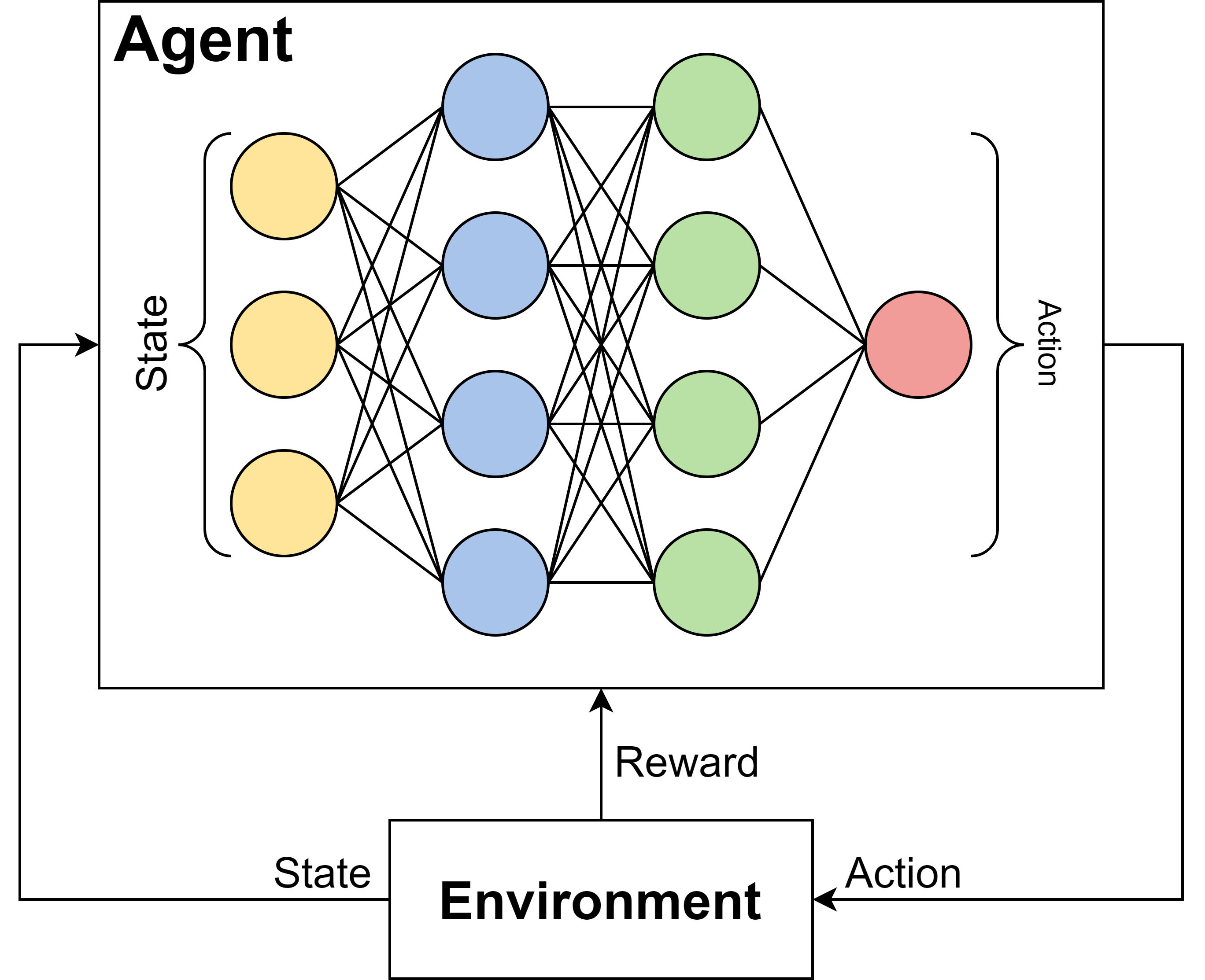}
    \caption{Architecture of Deep Reinforcement Learning.}
    \label{fig:drl_architecture}
\end{figure}

For RL problems with small discrete state-actions spaces, the state-action mapping can be stored in a table to approximate the mapping within a reasonable error value. However, for problems with large state-actions spaces, it is difficult to store such large amounts of data and, therefore, traditional RL methods suffer in terms of memory and performance. To overcome this, we can incorporate DRL, which is a combination of RL and deep neural networks. A neural network can be used to approximate a value or policy function. Essentially, neural nets learn to map states to values rather than using a lookup table. Thus, a DRL model can independently learn to establish a successful function for gaining maximum long-term rewards in RL problems with large state-actions spaces.

We have defined some characteristics within our DRL model in order for it to act as both a binary and multiclass classifier. For binary classification, we have defined our action space as follows:
\begin{itemize}
    \item 0 : No Alert (benign)
    \item 1 : Alert (attack)
\end{itemize}
And the rewards for this model are defined by:
\begin{itemize}
    \item +1 if agent correctly alerts to the correct type of attack.
    \item 0 if agent does not raise an alert when it is not needed.
    \item -1 if agent does not raise an alert when there is an attack.
    \item -1 if agent raises alert when there is not one needed.
\end{itemize}

For multiclass classification, we have defined our action space, also seen in Fig.~\ref{fig:drl_architecture}, as follows:
\begin{itemize}
    \item 0 : No Alert (benign)
    \item 1 : DoS
    \item 2 : Probe 
    \item 3 : R2L
    \item 4 : U2R
\end{itemize}
And the rewards for this model are defined by:
\begin{itemize}
    \item +1 if agent correctly alerts to the correct type of attack.
    \item 0 if agent does not raise an alert when it is not needed.
    \item -1 if agent does not raise an alert when there is an attack.
    \item -1 if agent raises alert when there is not one needed.
    \item -1 if agent raises an alert to the incorrect type of attack.
\end{itemize}

In terms of network security, alerting benign network traffic is typically safer than not alerting to an actual attack. Thus, we might consider that the reward for the latter two cases in the above enumeration should be greater than $-1$. However, this reward function was selected because identifying the wrong type of attack would lead to misdirection of resources, which we want to avoid.

Finally, the state space for both binary and multiclass classification is a collection of 41 features, both numerical and nominal, existing within the NSL-KDD dataset. Thus, we have an fairly complex and detailed state space. A visual of this environment can be seen in Fig.~\ref{fig:drl_architecture}.

\begin{figure}[t]
    \centering
    \includegraphics[width=\columnwidth]{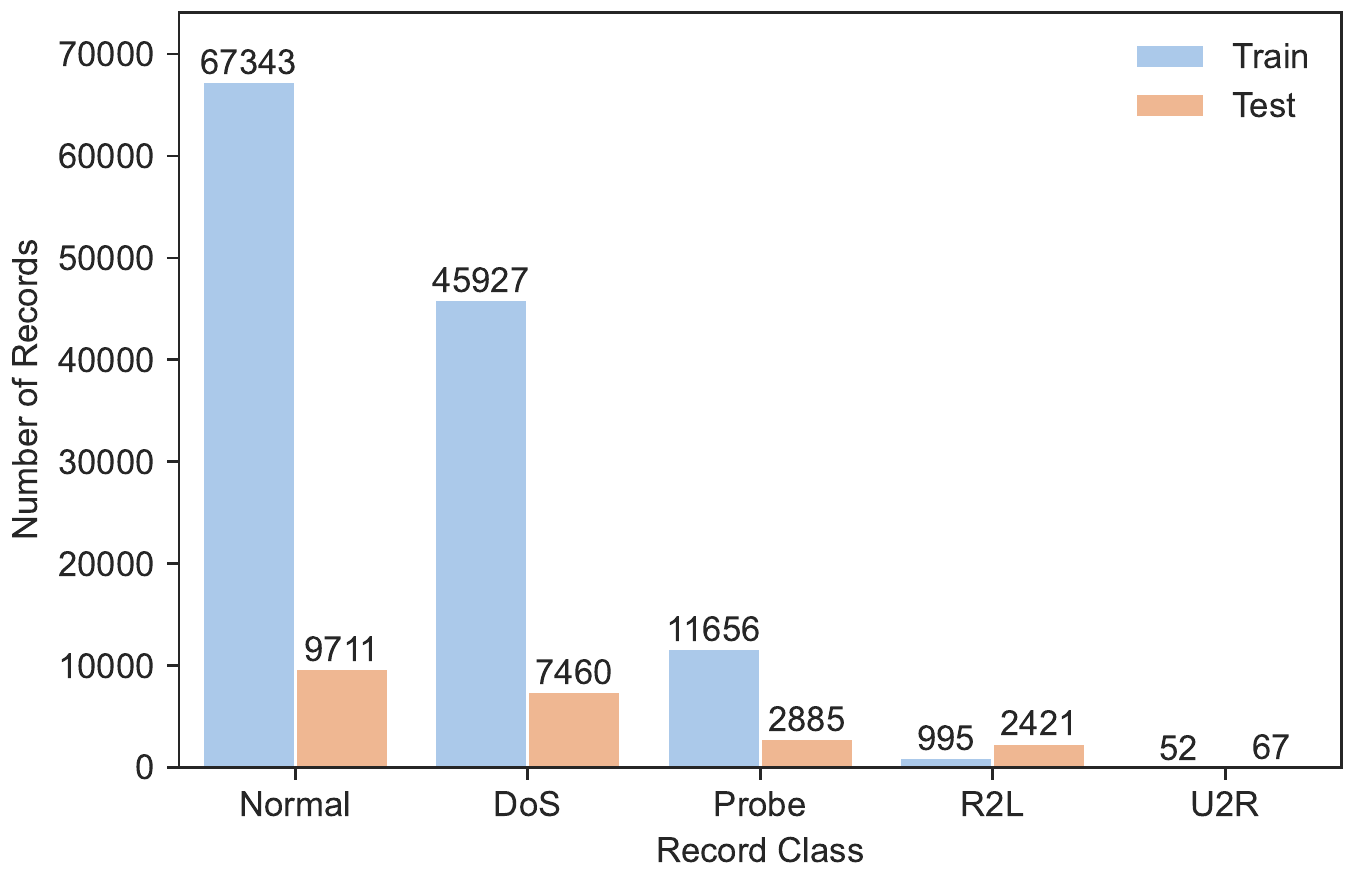}
    \caption{Distribution of NSL-KDD dataset by record classes.}
    \label{fig:nsl_kdd_distribution}
\end{figure}

\begin{figure}[t]
    \centering
    \includegraphics[width=\columnwidth]{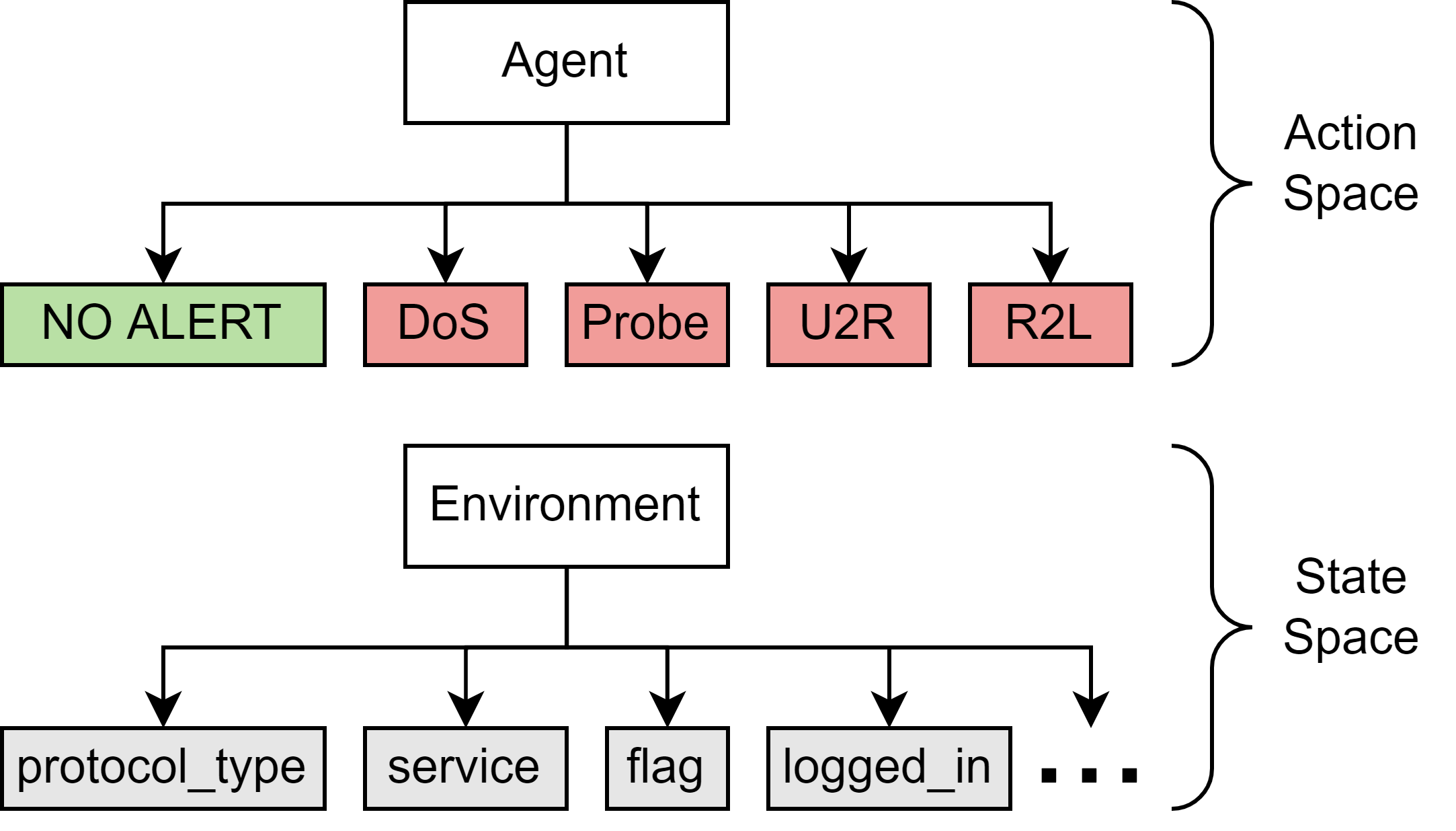}
    \caption{The action and state spaces of the proposed deep reinforcement learning model.}
    \label{fig:drl_action_state_space}
\end{figure}
 
\vspace{7mm}

In addition, we have assigned two distinct conditions for terminating an episode. An episode will be terminated if 1) it reaches a set timestep threshold, or 2) if an attack is issued and no alert has been made.


\section{Results}
\label{sec:results}

\begin{table}[t]
    \renewcommand{\arraystretch}{1.3}
    \centering
    \caption{Statistical Metrics for Synthetic Data}
    \label{tab:statistical_metrics}
    \begin{tabularx}{\columnwidth}{>{\centering\arraybackslash}m{0.4\columnwidth} YYY}
        \toprule
        \textbf{Synthetic Data} & \textbf{CSTest} & \textbf{KSTest} & \textbf{\thead{KSTest\\Extended}}\\
        \midrule
        CTGAN & 0.9971 & 0.9156 & 0.9181 \\
        CTGAN (Conditional) & 0.7468 & 0.8655 & 0.8571 \\
        CopulaGAN & 0.9988 & 0.9550 & 0.9574 \\
        CopulaGAN (Conditional) & 0.6982 & 0.9000 & 0.8864 \\
        \bottomrule
    \end{tabularx}
\end{table}

\begin{table}[t]
    \renewcommand{\arraystretch}{1.3}
    \centering
    \caption{Discrenment Results for Synthetic Data using\\Logistic Regression}
    \label{tab:detection_metrics}
    \begin{tabularx}{\columnwidth}{>{\centering\arraybackslash}m{0.4\columnwidth} Y}
        \toprule
        \textbf{Synthetic Data} & \textbf{Discernment Metric} \\
        \midrule
        CTGAN & 0.7579 \\
        CTGAN (Conditional) & 0.4139 \\
        CopulaGAN & 0.6862 \\
        CopulaGAN (Conditional) & 0.3948 \\
        \bottomrule
    \end{tabularx}
\end{table}

\begin{table*}[t]
    \renewcommand{\arraystretch}{1.3}
    \centering
    \caption{Machine Learning Performance for Binary Classification}
    \label{tab:binary_ml_models}
    \begin{tabularx}{\linewidth}{>{\centering\arraybackslash}m{0.2\linewidth} *{10}{c}}
        \toprule
        \multirow{2}[4]{*}{\textbf{Training Data}} &
        \multicolumn{2}{Y}{\textbf{Decision Tree}} &
        \multicolumn{2}{Y}{\textbf{\thead{AdaBoost\\Classifier}}} &
        \multicolumn{2}{Y}{\textbf{\thead{Logistic Regression\\Classifier}}} &
        \multicolumn{2}{Y}{\textbf{MLP Classifier}} &
        \multicolumn{2}{Y}{\textbf{Proposed DRL}} \\
        \cmidrule(lr){2-3}
        \cmidrule(lr){4-5}
        \cmidrule(lr){6-7}
        \cmidrule(lr){8-9}
        \cmidrule(lr){10-11}
        & Accuracy & F1 & Accuracy & F1 & Accuracy & F1 & Accuracy & F1 & Accuracy & F1 \\
        \midrule
        NSL-KDD & 0.8407 & 0.8414 & 0.8221 & 0.8270 & 0.8700 & 0.8802 & 0.8054 & 0.8080 & 0.8951 & 0.9064\\
        CTGAN & 0.8074 & 0.8112 & 0.8404 & 0.8486 & 0.8610 & 0.8710 & 0.8461 & 0.8545 & 0.8572 & 0.8687 \\
        CTGAN (Conditional) & 0.8801 & 0.8927 & 0.9086 & 0.9226 & 0.8740 & 0.8853 & 0.9077 & 0.9220 & 0.4662 & 0.1172\\
        CopulaGAN & 0.7735 & 0.7607 & 0.8259 & 0.8246 & 0.8163 & 0.8201 & 0.7918 & 0.7831 & 0.8294 & 0.8375\\
        CopulaGAN (Conditional) & 0.8287 & 0.8333 & 0.8743 & 0.8881 & 0.8256 & 0.8311 & 0.8947 & 0.9074 & 0.4901 & 0.1893\\
        \bottomrule
    \end{tabularx}
\end{table*}

\begin{table*}[t]
    \renewcommand{\arraystretch}{1.3}
    \centering
    \caption{Machine Learning Performance for Multi-label Classification}
    \label{tab:multiclass_ml_models}
    \begin{tabularx}{\linewidth}{>{\centering\arraybackslash}m{0.2\linewidth} *{9}{c}}
        \toprule
        \multirow{2}[2]{*}{\textbf{Training Data}} &
        \multicolumn{3}{Y}{\textbf{Decision Tree}} &
        \multicolumn{3}{Y}{\textbf{MLP Classifier}} &
        \multicolumn{3}{Y}{\textbf{Proposed DRL}} \\
        \cmidrule(lr){2-4}
        \cmidrule(lr){5-7}
        \cmidrule(lr){8-10}
        & Accuracy & F1 & F1 (weighted) & Accuracy & F1 & F1 (weighted) & Accuracy & F1 & F1 (weighted) \\
        \midrule
        NSL-KDD & 0.7685 & 0.5585 & 0.7338 & 0.7856 & 0.6302 & 0.7556 & 0.7300 & 0.4880 & 0.6891\\
        CTGAN & 0.7475 & 0.5297 & 0.7336 & 0.7765 & 0.6467 & 0.7572 & 0.4247 & 0.3033 & 0.4503\\
        CTGAN (Conditional) & 0.6200 & 0.4475 & 0.6525 & 0.7442 & 0.5643 & 0.7791 & 0.5520 & 0.3938 & 0.4533\\
        CopulaGAN & 0.7031 & 0.4165 & 0.6618 & 0.7374 & 0.4606 & 0.6863 & 0.7023 & 0.3967 & 0.6345\\
        CopulaGAN (Conditional) & 0.6116 & 0.3810 & 0.6215 & 0.7088 & 0.4401 & 0.6883 & 0.4839 & 0.2716 & 0.4049\\
        \bottomrule
    \end{tabularx}
\end{table*}

The following subsections describe the experimental results from our proposed GAN and DRL models, followed by a comparative analysis of our proposed model with other state-of-the-art ML methods.

\subsection{GAN Models}

For our experiments, we trained two GAN models, CTGAN~\cite{Xu2019} and CopulaGAN~\cite{SDVCopulaGAN}, using the implementations provided by the SDV open-source library \cite{montanez2018sdv}, following work by~\cite{Bourou2021} which showed promising results for generating network traffic data using models from this library. These models were trained on the NSL-KDD training data for 100 epochs with a batch size of 500. Both models used discriminator steps of 5, matching WGAN~\cite{Arjovsky2017}, an extended version of vanilla GAN. For the other hyperparameters, we opted to use the defaults provided by the SDV library.

The trained CTGAN and CopulaGAN models were each used to generate two synthetic datasets:
\begin{enumerate}
    \item A dataset containing 200\,000 records generated through regular sampling without conditions. As expected, these datasets contained records that closely matched the imbalanced class distribution of the original NSL-KDD dataset.
    \item A dataset containing 20\,000 records for each class generated using conditional sampling through rejection. These datasets were used to explore the efficacy of using GANs to generate a balanced distribution from an highly imbalanced training distribution.
\end{enumerate}

The statistical metric results showcased in Table~\ref{tab:statistical_metrics} indicate that both CTGAN and CopulaGAN model the discrete and continuous features of the NSL-KDD dataset effectively. As dictated by the KSTest and KSTestExtended, CopulaGAN models continuous features better than CTGAN and maintains parity for discrete features as indicated by the CSTest. Table~\ref{tab:detection_metrics} highlights the results for a linear regression classifier used to evaluate detection performance of the synthetic data. Altogether, the classifier found it challenging to distinguish the synthetic records from the real ones, which indicates that the GANs are able to capture aspects of the true dataset. Table~\ref{tab:binary_ml_models} and Table~\ref{tab:multiclass_ml_models} showcase the performance of ML models when trained to distinguish between various real and synthetic datasets. Across the board, there is comparable performance between the original real NSL-KDD dataset and the CTGAN and CopulaGAN synthetic datasets. Thus, there is promise in using synthetic data in place of real data.


\subsection{DRL Model}

The DRL models were implemented using both OpenAI Gym~\cite{Brockman2016} and Tensorflow~\cite{Abadi2016}. Training of the model was done in two distinct stages to investigate the variation in performance -- binary classification and multiclass classification.

\subsubsection{Binary Classification}

We begin with binary classification, using an action space of two (`alert' or `no alert'). While binary classification offers the user less knowledge on attack type specifications, it should perform the basic task of an IDS -- alerting the user to an attack with a high accuracy.

Initially, we trained the DRL model on the NSL-KDD training set, described in detail above. We did this to create a baseline to see how well our synthetic GAN-generated data performed in comparison. Prior to training our model, we converted all class labels using a binary mapping. If the class was originally `normal', we assigned it a value of `0', otherwise it was assigned a value of `1', implying that the data point was an attack of some sort.

For each model, proximal policy optimization (PPO2), a policy-gradient algorithm that directly optimizes the expected reward by estimating the gradient of the policy from the trajectories taken by the agent, is executed. We applied a custom multi-layer perceptron, a class of feedforward neural network \cite{noriega2005multilayer}, of three layers with size 128, 64, and 32. In addition, each model used a rectified linear unit (ReLU) activation function.

\begin{figure}[t]
    \centering
    \includegraphics[width=0.95\columnwidth]{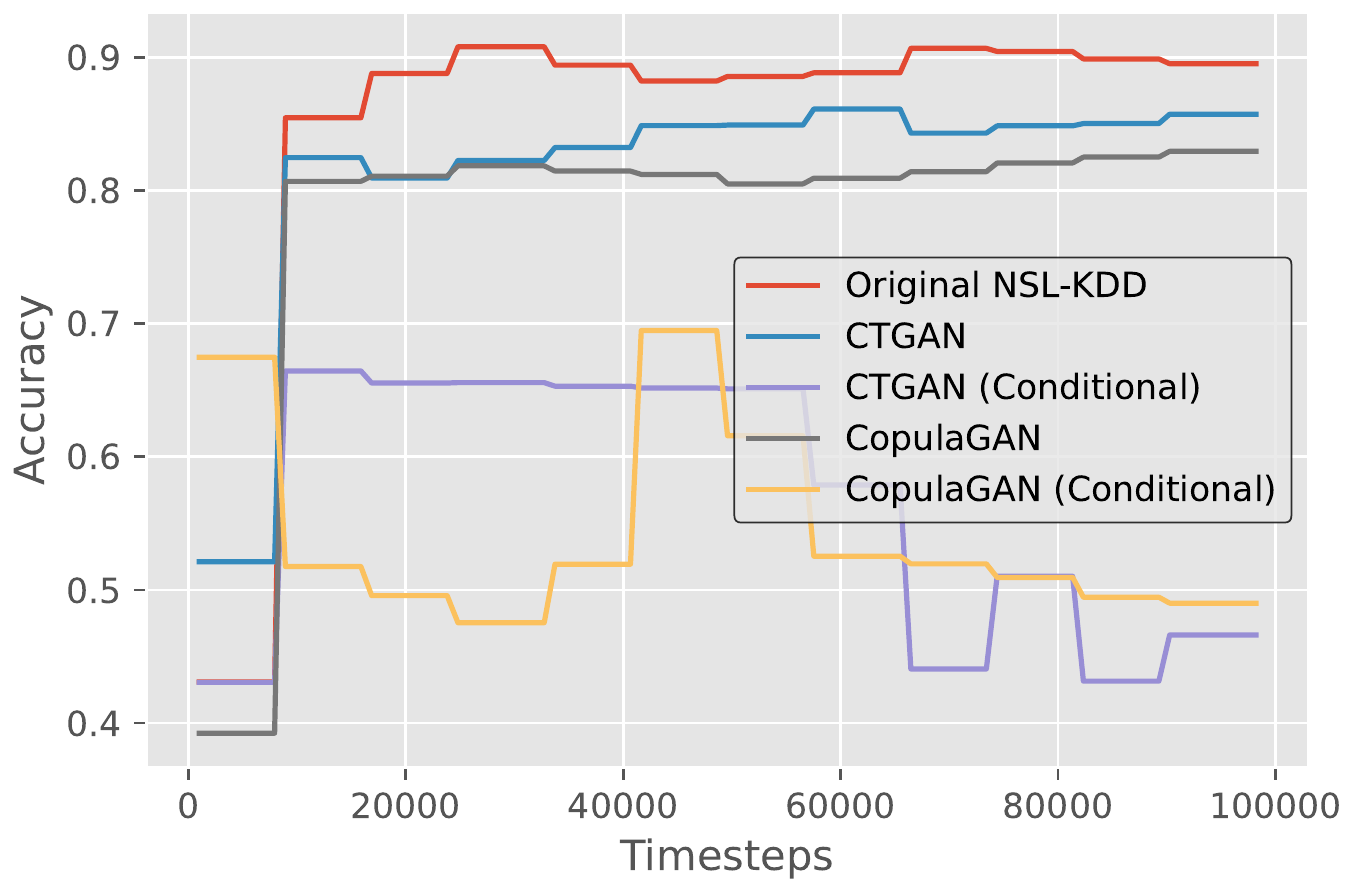}
    \caption{Results measuring the accuracy of binary classification after training the DRL model on both the original NSL-KDD dataset and each of the synthetic GAN datasets.}
    \label{fig:binomial_classifier_accuracy}
\end{figure}

Training on the NSL-KDD training dataset for 100,000 timesteps resulted in an average accuracy of \textit{89.5\%} and an F1-score of \textit{0.906} on the test dataset. We then proceeded to train the DRL model on each of the GAN-generated datasets one-by-one and evaluate them individually on the NSL-KDD test dataset. The detailed results of these experiments can be seen in Table \ref{tab:binary_ml_models}, and viewed in terms of progressive performance for average accuracy in Fig.~\ref{fig:binomial_classifier_accuracy}.

Training on CTGAN synthetic data performs the best after the NSL-KDD trained model, with \textit{85.7\%} accuracy and 0.869 F1-score. Training using CopulaGAN synthetic data trails close behind with \textit{82.9\%} accuracy and 0.838 F1-score. The conditional variations of both CopulaGAN and CTGAN perform significantly worse than the three other datasets, reaching their peak of \textit{70\%} and \textit{66\%} respectively almost immediately and then dropping to just below \textit{50\%}.

\subsubsection{Multiclass Classification}

\begin{figure*}[t]
    \centering
    \subfloat[NSL-KDD]{\includegraphics[width=0.3\linewidth]{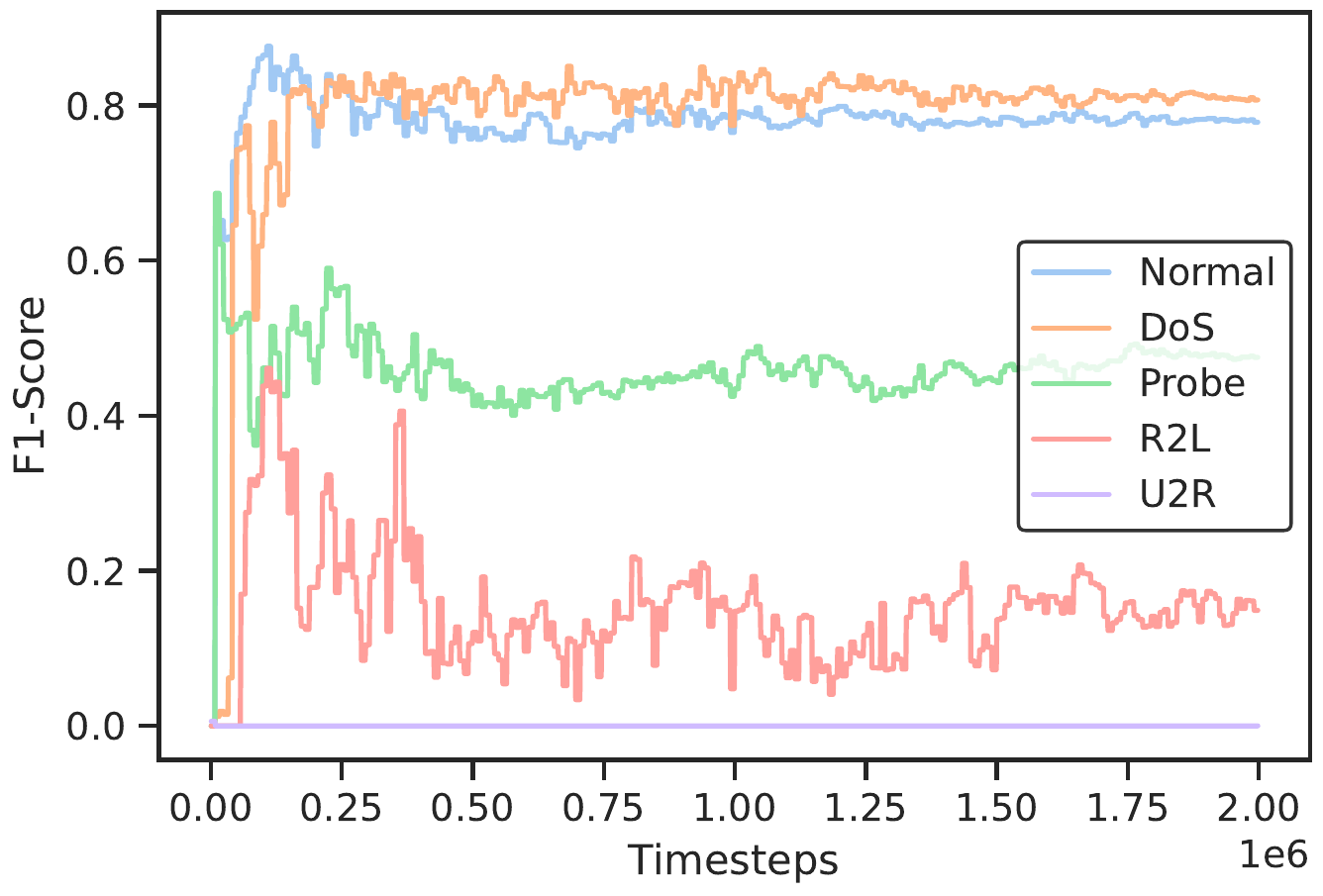}%
    \label{fig:multinomial_results_a}}
    \hfil
    \subfloat[CTGAN]{\includegraphics[width=0.3\linewidth]{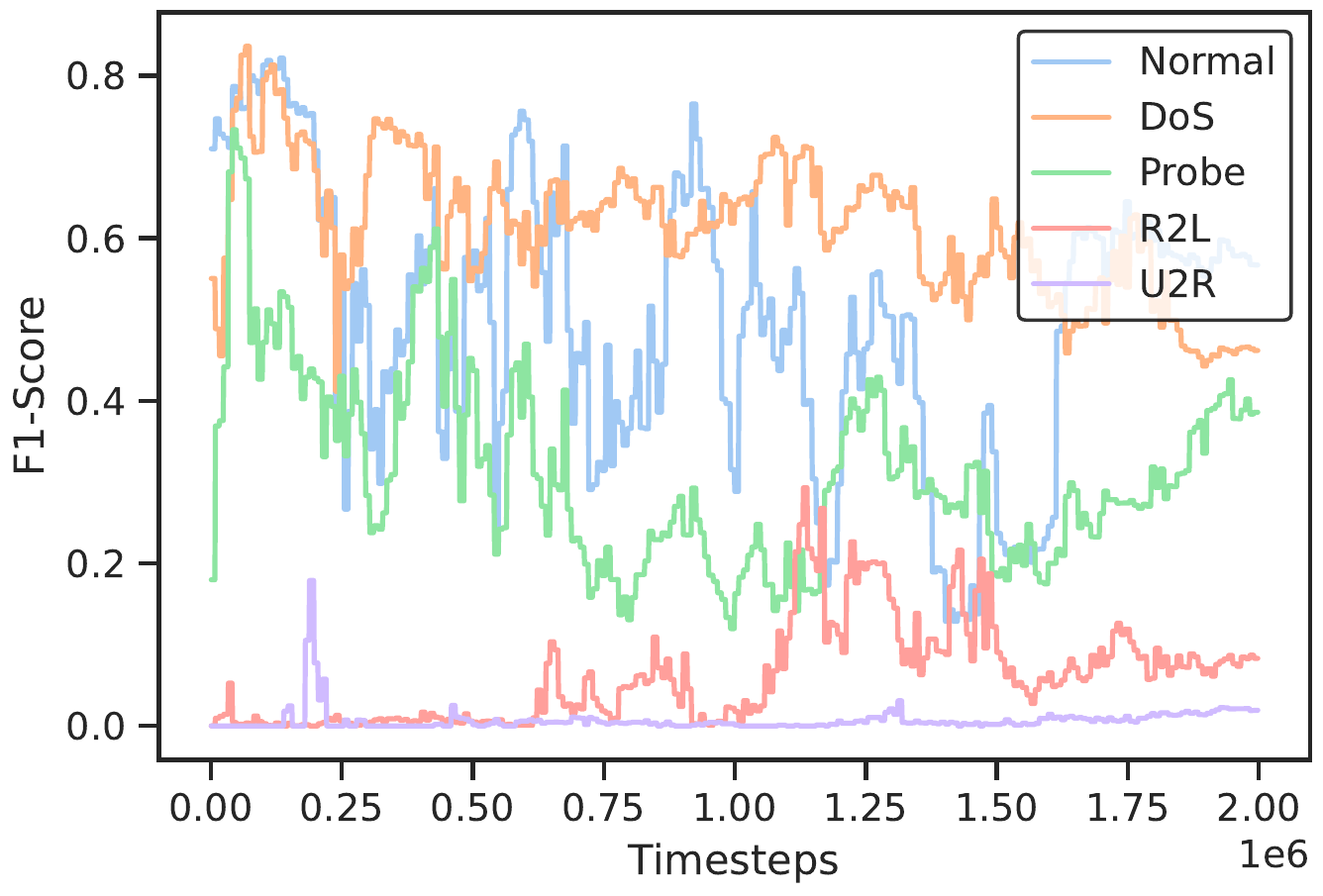}%
    \label{fig:multinomial_results_b}}
    \hfil
    \subfloat[CTGAN (Conditional)]{\includegraphics[width=0.3\linewidth]{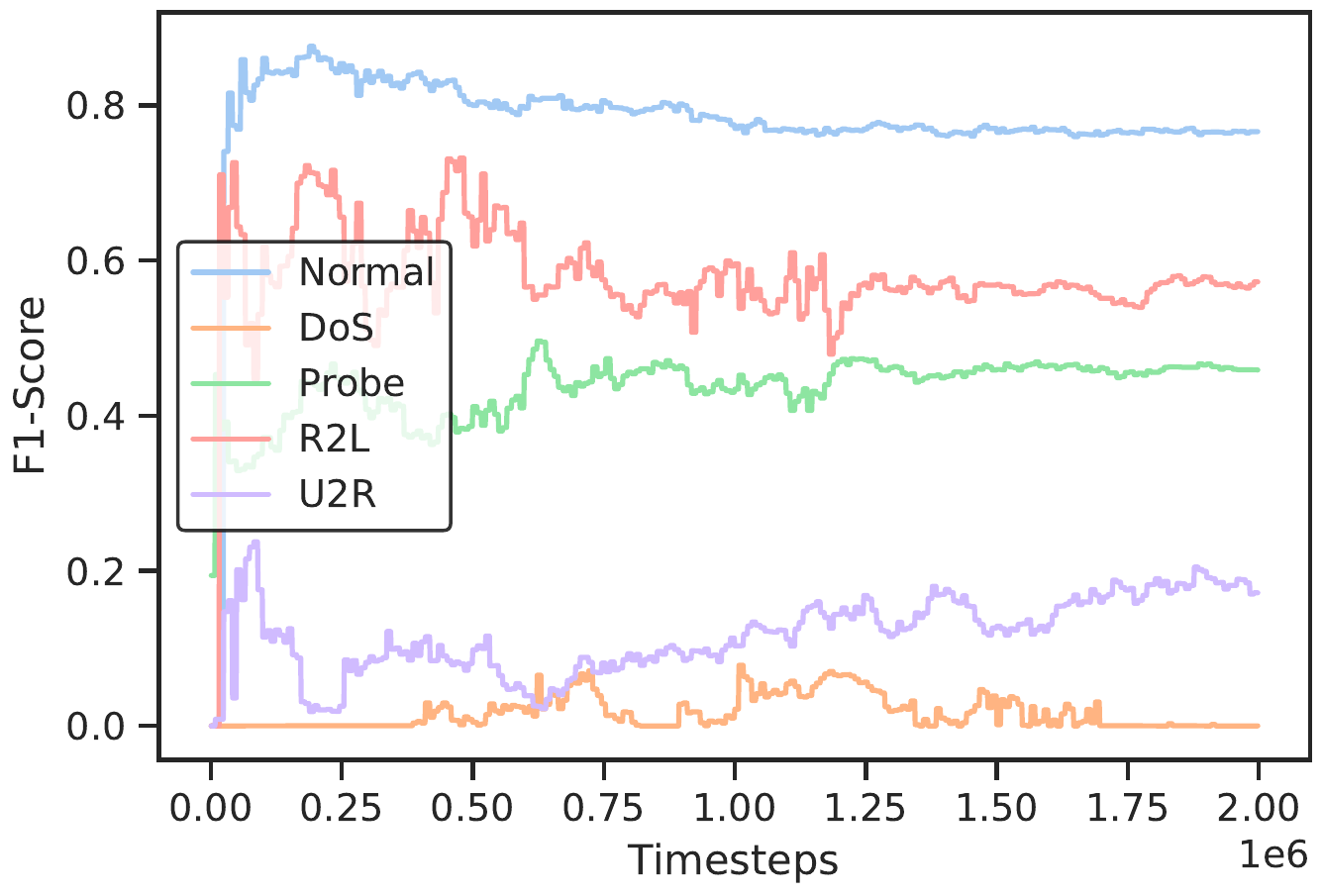}%
    \label{fig:multinomial_results_c}}
    \hfil
    \subfloat[CopulaGAN]{\includegraphics[width=0.3\linewidth]{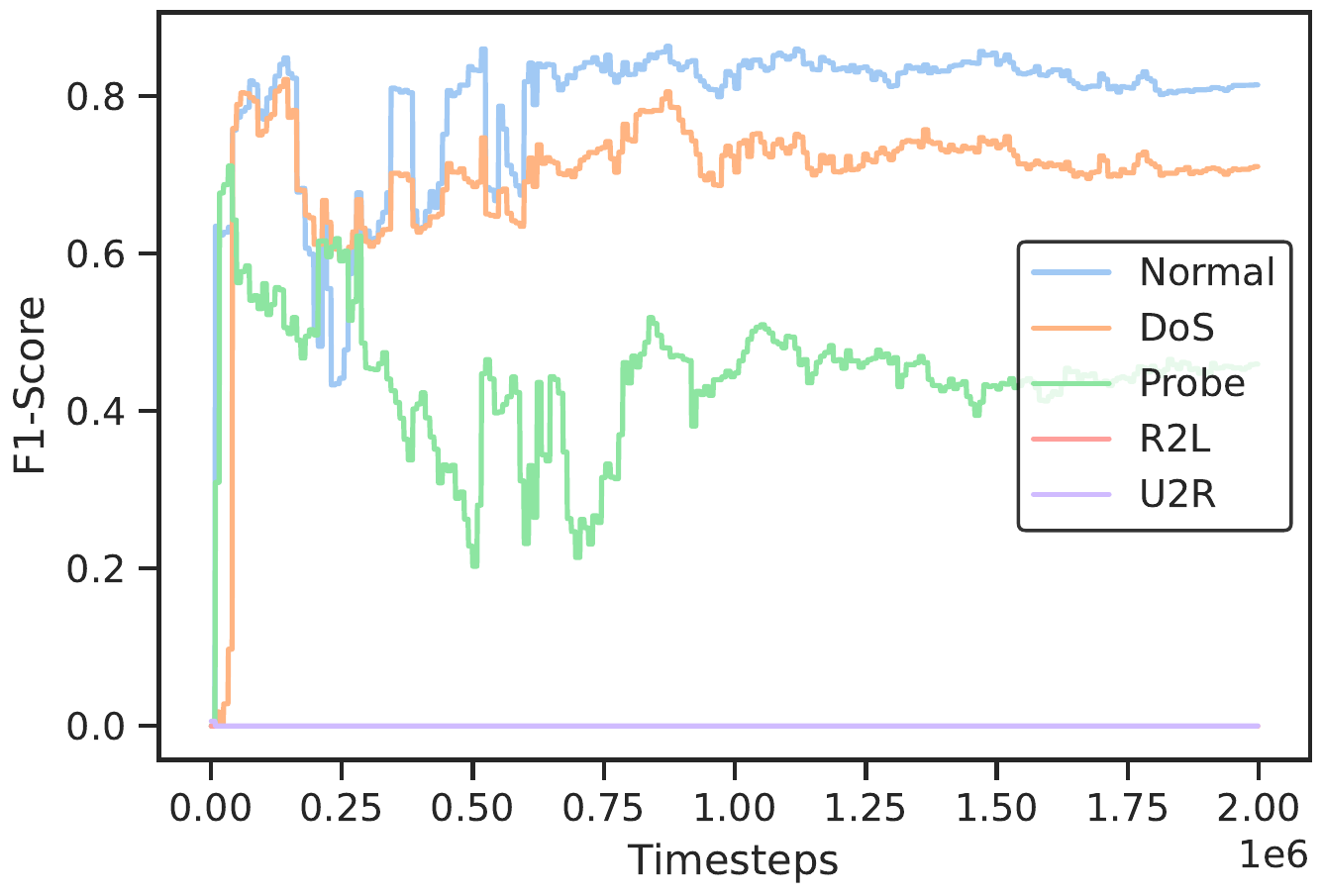}%
    \label{fig:multinomial_results_d}}
    \hfil
    \subfloat[CopulaGAN (Conditional)]{\includegraphics[width=0.3\linewidth]{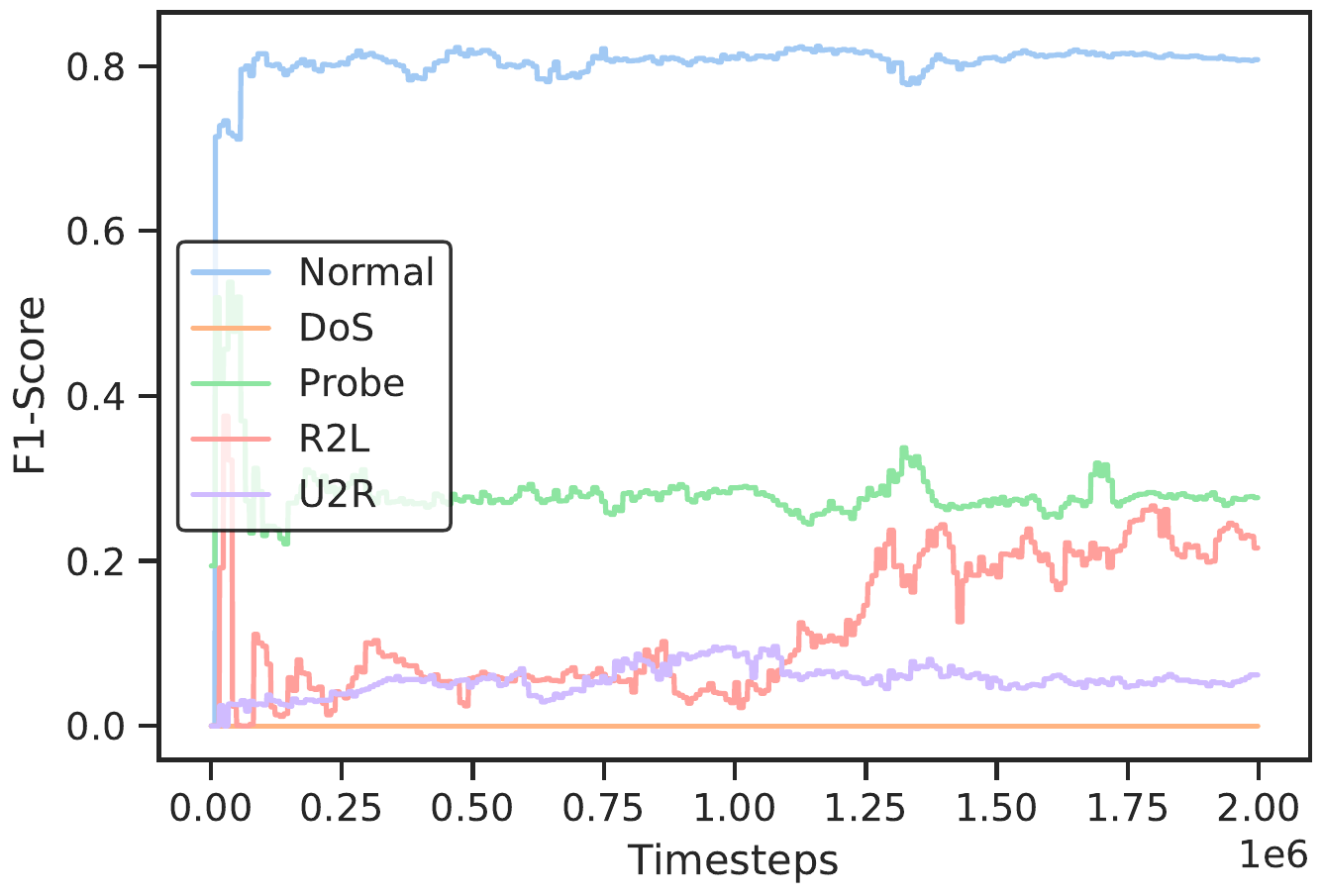}%
    \label{fig:multinomial_results_e}}
    \hfil
    \subfloat[Average Accuracy]{\includegraphics[width=0.3\linewidth]{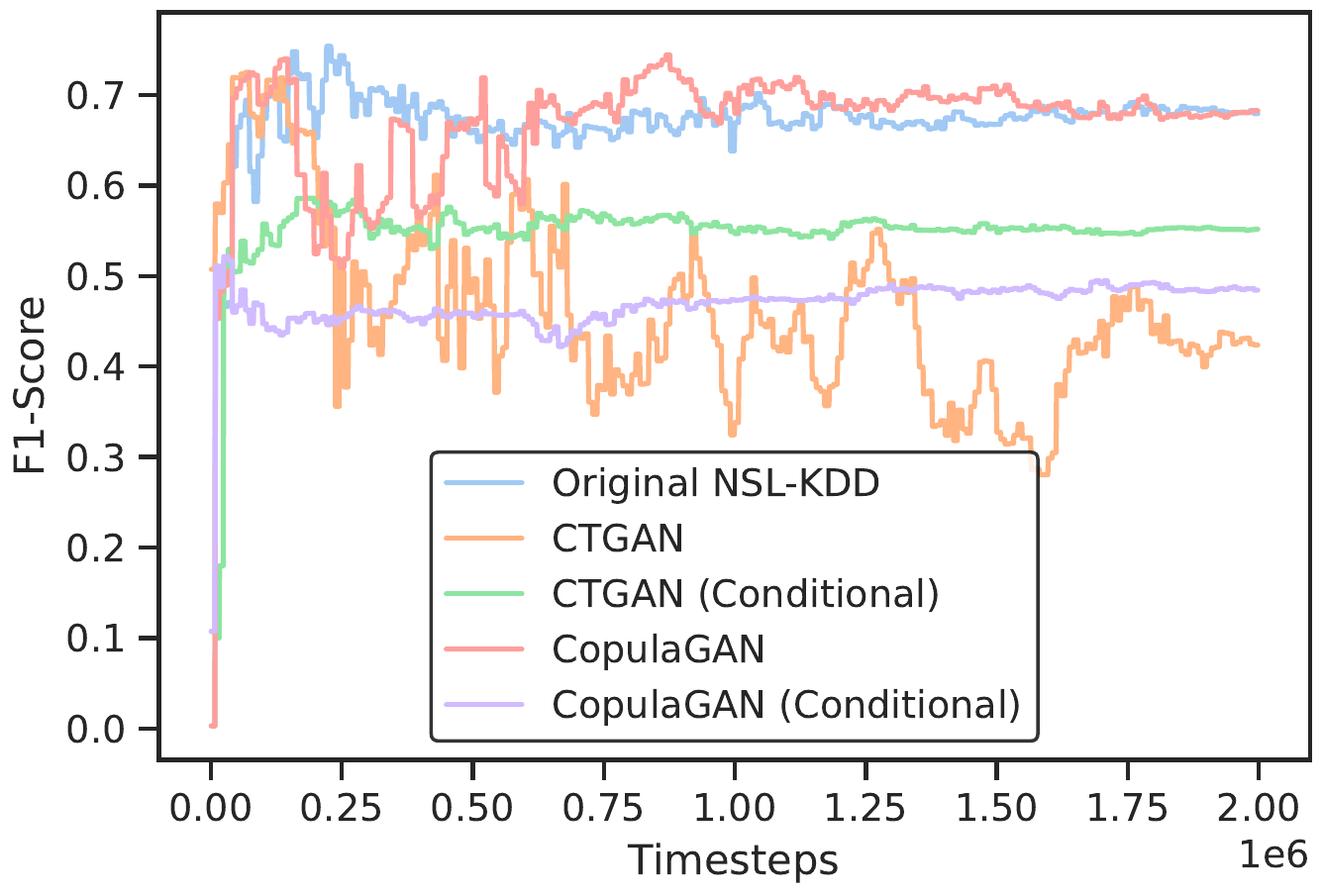}%
    \label{fig:multinomial_results_f}}
    \caption{Results measuring the F1-scores of multiclass classification (fig a-e) as well as the averages (fig f) after training the DRL model for 2 million timesteps on both NSL-KDD as well as each synthetic dataset.}
    \label{fig:multinomial_results}
\end{figure*}

\begin{table}[t]

    \renewcommand{\arraystretch}{1.3}
    \centering
    \caption{Class-Based F1 Scores for Multi-Label Classification}
    \label{tab:multinomial_f1_scores}
    \begin{tabularx}{\linewidth}{>{\centering\arraybackslash}m{0.2\linewidth} *{9}{c}}
        \toprule
        \textbf{Dataset} & \textbf{Normal} & \textbf{DoS} & \textbf{Probe} & \textbf{R2L} & \textbf{U2R}\\
        \midrule
        NSL-KDD & 0.7785 & 0.8072 & 0.4752 & 0.1490 & 0.0 \\
        CTGAN & 0.5670 & 0.4618 & 0.3858 & 0.0831 & 0.0192\\
        CTGAN (Conditional) & 0.7662 & 0.0 & 0.4589 & 0.5725 & 0.1716 \\
        CopulaGAN & 0.8139 & 0.7101 & 0.4593 & 0.0 & 0.0\\
        CopulaGAN (Conditional) & 0.8039 & 0.0 & 0.2201 & 0.2097 & 0.0512 \\
        \bottomrule
    \end{tabularx}
\end{table}

We then trained the DRL models to perform multiclass classification. Similar to binary classification, we are still detecting whether there is an attack or not, however we now attempt to classify which type of attack is taking place. Instead of `0' or `1', our action space consists of 0, 1, 2, 3, and 4. As stated previously, 0 maps to `benign', whereas 1, 2, 3, and 4 map to DoS, Probe, R2L, and U2R respectively. As our action space has increased in comparison to binary classification, our problem becomes significantly larger and more challenging.

Like binary classification, we used a ReLU activation function, however for the conditional versions of both CTGAN and copulaGAN we used a Sigmoid activation function, as we found that this results in a significant increase in performance on test data. For each model, we again used a custom multi-layer perceptron of three layers with size 128, 64, and 32.

Again, we first analyzed the performance of our model after being trained on the real NSL-KDD dataset in order to create a benchmark. As seen in Table \ref{tab:multiclass_ml_models}, our DRL model achieved \textit{73\%} accuracy and a \textit{68.9\%} weighted F1-score.

We then trained the DRL model on the four GAN-generated synthetic datasets discussed previously. The most promising results were seen in training the model on CopulaGAN. The model reaches an accuracy of \textit{70.2\%} and a weighted F1-score of \textit{63\%}. This is just a \textit{2.7\%} drop in accuracy from training on the true NSL-KDD data. Training the DRL model on the remaining three synthetic datasets underperforms when compared to both the decision tree and MLP classifier.

As discussed previously, an F1-score refers to both precision and recall being high. When we train on imbalanced datasets, the F1-scores in minority classes are typically quite low, as the ML model does a poor job of recognizing and properly classifying that test data. Looking at Table \ref{tab:multinomial_f1_scores}, we can see the F1-scores for each individual class for each of our training sets. Since NSL-KDD had extremely low records for both R2L and U2R, we can see that the F1-scores for these classes are also quite low at \textit{0.1490} and \textit{0.0}, respectively.

One of the major goals of our work was to determine if, by generating synthetic GAN data, we could inflate the F1-scores (more specifically, precision and recall) of the minority classes from our imbalanced dataset. In Table \ref{tab:multinomial_f1_scores}, we can see that training our DRL model with data generated from conditional CTGAN and conditional CopulaGAN improved upon the F1-scores for both R2L and U2R in the same way that we would expect to see if the true dataset naturally contained more records of these two class types. Training the DRL model on synthetic data from conditional CTGAN increased the F1-scores for R2L and U2R by 0.573 and 0.172 respectively. Training on synthetic data from conditional CopulaGAN improved the F1-scores for R2L and U2R by 0.210 and 0.051 respectively. This demonstrates that the concept of using GAN models to generate synthetic data for a minority class and artificially inflating the training set in order to have better performance in classifying underrepresented classes is a viable option.




%

\section{Conclusion}
\label{sec:conclusion}

In this paper, we have proposed a SNIDS which is able to perform binary and multiclass classification on network traffic data. We used DRL to implement this IDS. The model was trained using the NSL-KDD dataset, allowing it to detect a range of attack types on a network. To enhance the learning capabilities of our proposed model, GANs were used to fabricate training data. Our results demonstrate that this system is able to interact with the network and identify attack classes with competitive accuracy. As well, we show that generating synthetic data for underrepresented classes can improve the precision and recall within these classes, thus acting as a solution for imbalanced datasets.

For binary classification, we obtained an \textit{89.5\%} accuracy after training on the NLS-KDD dataset. We consider this our baseline model. When trained on the four synthetic datasets, data generated from unconditional CTGAN produced an accuracy of \textit{85.7\%}, the closest competition to the baseline model.

For multiclass classification, we obtained a \textit{73.0\%} accuracy after training on the NSL-KDD dataset. When trained on the four synthetic datasets, data generated from CopulaGAN produced an accuracy of \textit{70.2\%}, the closest competition to the baseline model. Thus, clearly our GAN models generate data realistic enough to create competitive IDS.

Further, both Table \ref{tab:multinomial_f1_scores} and Fig.~\ref{fig:multinomial_results} demonstrate an increase in F1-scores for minority classes on the IDS trained using GAN-generated data. Thus, while our overall accuracy decreased, we are getting better precision and recall performance for the classes without sufficient data in the NSL-KDD dataset. This points to a solution for other ML models trying to learn from imbalanced datasets.

\section{Future Work}
\label{sec:future_work}

While our work demonstrated competitive classifiers and an increase in individual F1-scores for minority classes, there is still room for improvement.

When training our GAN models, we have discussed that we used 41 features from the NSL-KDD dataset as input to our model. There are two major changes that we aim to implement in our future work. First, passing our input dataset through a pipeline of feature analysis methods, including (but not limited to) Pearson correlation, recursive feature elimination, and Lasso, with the aim to reduce our feature space. This has the potential to increase the quality of our generated dataset, thus increasing the evaluation metric scores for our DRL model. Secondly, supplementing the NSL-KDD dataset with data from under-represented classes in order to balance the dataset. Our work demonstrates that there is a notable increase in F1-score when the class has a significant amount of data being given as input to the GAN model. We plan to explore this idea, and see the limitations of our performance when the GAN is trained on significant sample sizes from each class, rather than just a small subset.

We also plan to explore GAN models that are trained only on the minority classes of our true dataset classes. Thus, this generated data could potentially be merged with the true dataset to allow for heightened overall performance of the IDS, as we are synthetically creating balance.

Finally, we plan to explore the performance of training both DQN, a value-iteration based method \cite{yoon2017deep}, and A3C \cite{babaeizadeh2016ga3c}, a value-iteration and policy-gradient method, on GAN-generated data to see how it compares with our PPO2 model. Both DQN and A3C are common DRL approaches, and have the potential to surpass the performance of our current model.

\bibliographystyle{IEEEtran}
\bibliography{IEEEabrv,main}

\vskip -2.5\baselineskip plus -1fil 
\begin{IEEEbiographynophoto}{Caroline Strickland}
received a B.Sc in 2017 and M.Sc in 2019 from Memorial University of Newfoundland. She is currently in the third year of pursuing a PhD in Computer Science at the University of Western Ontario. Her past research has focused on using reinforcement learning for pattern formation within swarm systems, and her current research interests involve the intersection of hierarchical reinforcement learning with healthcare.
\end{IEEEbiographynophoto}
\vskip -2.5\baselineskip plus -1fil 
\begin{IEEEbiographynophoto}{Chandrika Saha}
received a B.Sc in Computer Science and Engineering from the University of Barishal,  in 2019. Currently, she is pursuing her M.Sc. in Computer Science at the Western University, London, Ontario, Canada. Her research interest is Machine Learning, more specifically, deep learning and its application to network security. 
\end{IEEEbiographynophoto}
\vskip -2.5\baselineskip plus -1fil 
\begin{IEEEbiographynophoto}{Muhammad Zakar}
received a B.Sc. in Computer Science from Western University, London, Canada, in 2021. Currently, he is pursuing a M.Sc. in Computer Science at Western University. His current research interests are in the areas of autonomous drones and vehicles, distributed systems, next-generation networks, and machine learning.
\end{IEEEbiographynophoto}
\vskip -2.5\baselineskip plus -1fil 
\begin{IEEEbiographynophoto}{Sareh Soltani Nejad}
received a B.Sc. in Computer Engineering from Amirkabir University of Technology (AUT), Tehran, Iran, in 2019. She is currently pursuing a M.Sc. in Computer Science at the University of Western Ontario, Canada. Her research interests broadly focus on Machine learning and Internet of Things applications in Smart Cities, Smart homes and Healthcare.
\end{IEEEbiographynophoto}
\vskip -2.5\baselineskip plus -1fil 
\begin{IEEEbiographynophoto}{Noshin Tasnim}
received a B.Sc. in Computer Science and Engineering from BRAC University, Bangladesh, in 2019. She is currently pursuing a M.Sc. in Computer Science with the Department of Computer Science, Western University, London, Canada. Her current research interests are in the areas of network security, and machine learning.
\end{IEEEbiographynophoto}
\vskip -2.5\baselineskip plus -1fil 
\begin{IEEEbiographynophoto}{Daniel Lizotte}
is currently an Associate Professor in the Department of Computer Science and in the Department of Epidemiology and Biostatistics, University of Western Ontario, London, ON, Canada. His research group in collaboration with community partners investigates different aspects of data-driven decision support in public health and health care. This work aligns with methodological research in the areas of machine learning, epidemiology, and biostatistics. He has received funding from the Natural Sciences and Engineering Research Council of Canada, the Canadian Institutes of Health Research, and the Social Sciences and Humanities Research Council of Canada, and he has served in various capacities on committees for the Machine Learning for Health Care, International Conference on Machine Learning, and NeurIPS conferences.
\end{IEEEbiographynophoto}
\vskip -2.5\baselineskip plus -1fil 
\begin{IEEEbiographynophoto}{Anwar Haque}
is an Assistant Professor in the Deptartment of Computer Science at the University of Western Ontario, Canada. Before joining Western, he was an Associate Director at Bell Canada. He is a leading international expert on next-generation communication network resources and performance management, cyber security, and smart city applications. Dr. Haque has authored/co-authored over 80 peer-reviewed research publications in leading journals and conferences, authored many industry technical papers, and held a number of patent/licenses. He has been awarded several national/provincial-level research grants, including NSERC, MITACS, OCE, and SOSCIP. Dr. Haque's collaborative research grants are valued at more than \$15 million. Dr. Haque is serving on the inaugural advisory committee for the newly established Bell-Western 5G research centre, and he established an industry consortium to promote and support smart systems and digital services research at Western. Dr. Haque is the director of the Western Information \& Networking Group (WING) Lab at Western.
\end{IEEEbiographynophoto}

\end{document}